\documentclass[journal]{IEEEtran}%\IEEEoverridecommandlockouts

\usepackage[T1]{fontenc}
\DeclareSymbolFont{letters}{OML}{cmex}{m}{it}
\usepackage{filecontents}

\usepackage{epsfig}
\usepackage{graphicx}
\graphicspath{{./figures/}}
\usepackage{amstext}
\usepackage{setspace}
\usepackage{float}
\usepackage{amsmath,amssymb,amsthm,amsfonts}
\usepackage{cite}
\usepackage{latexsym}
\usepackage{dsfont}
\usepackage{epstopdf}
\usepackage{verbatim}
\usepackage{mathrsfs}
\usepackage{amsbsy}
\usepackage{amsxtra}
\usepackage{amsopn}
\usepackage{multirow}
\usepackage{tikz,makecell}
\usepackage{array}
\usepackage{subfigure}
\usepackage{psfrag}
\usepackage{nomencl}
\usepackage{lineno}
\usepackage{algorithm}
\usepackage{algorithmic}
\usepackage{pgfplots}
\usepackage{scalefnt}
\usetikzlibrary{decorations.pathreplacing,automata,calc,positioning,arrows}

\definecolor{aqua}{rgb}{0.0, 1.0, 1.0}

\makeindex

\newlength
\figureheight
\newlength
\figurewidth

\newcommand{\bigp}[2]{\left(}

\makeatletter
\def\thm@space@setup{\thm@preskip=0pt
\thm@postskip=0pt}
\makeatother
\newtheoremstyle{newstyle}
{2pt} %Aboveskip
{1pt} %Below skip
{} %Body font e.g.\mdseries,\bfseries,\scshape,\itshape
{12pt} %Indent
{\itshape} %\bfseries %Head font e.g.\bfseries,\scshape,\itshape
{:} %Punctuation after theorem header
{5pt} %Space after theorem header
{\thmname{#1}\thmnumber{ #2}\thmnote{(#3)}}

\theoremstyle{newstyle}

\theoremstyle{newstyle}
\newtheorem{remark}{Remark}

\theoremstyle{newstyle}
\usepackage{color,soul}

\usepackage{float}
\usepackage{diagbox}

\usepackage{enumitem}
\usepackage{tabularx,booktabs}
\newcolumntype{C}{>{\centering\arraybackslash}X} % centered version of "X" type
\usepackage{lipsum}

\newtheorem{theorem}{Theorem}
\newtheorem{lemma}{Lemma}

\newtheorem{corollary}{Corollary}

\begin{document}

\title{Distributed Join-the-Idle-Queue for Low Latency Cloud Services}

\author{\IEEEauthorblockN{Chunpu Wang, Chen Feng,~\IEEEmembership{Member,~IEEE} and Julian Cheng,~\IEEEmembership{Senior Member,~IEEE}}
% \IEEEauthorblockA{School of Engineering, University of British Columbia, Kelowna, Canada\\
% chunpu.wang@alumni.ubc.ca, chen.feng@ubc.ca, julian.cheng@ubc.ca}
% \thanks{M. Shell was with the Department
%of Electrical and Computer Engineering, Georgia Institute of Technology, Atlanta,
%GA, 30332 USA e-mail: (see http://www.michaelshell.org/contact.html).}% <-this % stops a space
\thanks{Chunpu Wang, Chen Feng and Julian Cheng are with School of Engineering, The University of British Columbia (Okanagan Campus), Kelowna, Canada. E-mail: chunpu.wang@alumni.ubc.ca, chen.feng@ubc.ca and julian.cheng@ubc.ca.}% <-this % stops a space
%\thanks{Manuscript received April 19, 2005; revised August 26, 2015.}

 }

\maketitle
%% =================================
\begin{abstract} 
Low latency is highly desirable for cloud services.
To achieve low response time, stringent timing requirements are needed for task scheduling
in a large-scale server farm spanning thousands of servers. In this paper, we conduct an in-depth analysis for distributed
Join-the-Idle-Queue (JIQ), a promising new approximation of an idealized task-scheduling algorithm.
In particular, we derive semi-closed form expressions for the delay performance of distributed JIQ,
and we propose a new variant of distributed JIQ that offers clear advantages over alternative algorithms
for large systems. 

% \textcolor{red}{review comment}

% \textcolor{blue}{close form formula}

% \textcolor{green}{writing standards}
\end{abstract}

\vspace{2mm}


\section{INTRODUCTION}

\subsection{Motivation}
%Nowadays, low latency is one of the most important standards for online applications and services. For instance, Google requires a search conducted in less than one second. For some real-time financial tools (e.g., Yahoo Finance) the latency requirement is even stricter. 
%Meanwhile, the daily usage of online applications is growing fast. 
%Take Facebook as an example. According to the Social Skinny, Facebook users posted 510 comments, 293,000 statuses and 136,000 photos for every 60 seconds in 2016. 
%Daily usage of online applications is so frequent that huge amount of tasks are created every second.
%Hence, more servers are required to cope with such massive incoming tasks.
%Even as of June 2014, the Facebook was running at roughly 180,000 servers in its datacenters.

In cloud communication, low latency is highly desirable for online services
spanning thousands of servers. For example, Google search typically returns the query results within a few hundreds of milliseconds.
According to Google and Amazon, an extra latency of $500$ milliseconds in response time could result in a $1.2\%$ loss of users and revenue \cite{Sch09}.
The demand for fast response time, which significantly impacts user experience and service-provider revenue, is translated into
stringent timing requirements for task scheduling in a large-scale server farm.

%Hence, more servers are required to cope with such massive incoming tasks.  Similarly, Amazon Web Services is likely to have 1.3 million servers in 2016.
Join-the-Shortest-Queue (JSQ) is an idealized algorithm to achieve short response time. It tracks the queue lengths of all the servers
and selects the least loaded server for a newly arrival task. Although JSQ is proven to be latency optimal \cite{Ery12}, it doesn't scale well
as the system size increases. The reason is that  tracking the global queue-length information is both time and resource consuming.
To alleviate this problem, the Power-of-$d$-Choices (Po$d$) algorithm ($d \ge 2$) has been proposed as an ``approximation" of the idealized JSQ.
Instead of tracking the global information, Po$d$ only probes $d$ servers uniformly at random upon a task arrival and 
selects the least loaded one for the new task. Although Po$d$ achieves reasonably good average response time \cite{Mit96},
its tail response time still remains high for large-scale systems \cite{Kay13, Lei15} and its probing operation incurs additional delay.

%The scheduler is an essential matchmaker between tasks and servers. It assigns tasks to suitable servers to make them finished (i.e., task response time) as soon as possible. 
%
%Ideally, the  scheduler tracks the status of all the servers to find out the least loaded server (i.e., the server with the shortest task queue) when a new task arrives. As the tracking procedure is both time and resource consuming, it only applied to small size systems. 
%
%In large scale system, the Power of $d$ Choices (Po$d$) is proposed to approximate the performance of JSQ. Instead of tracking the status of all the servers, the Po$d$ scheduler only probes $d$ servers uniformly at random at each task arrival. However, it could perform badly as lack of the information of all servers.

Recently, distributed Join-the-Idle-Queue (JIQ)  \cite{Lu11} has emerged as a promising new approximation of JSQ.
JIQ employs a number of distributed schedulers, each maintaining an I-queue that stores a list of idle servers.
When a new task arrives at the system, it randomly visits a scheduler\textcolor{red}{,} asking to join an idle server in its I-queue.
Compared to JSQ, each scheduler in JIQ only maintains local information and is scalable to large systems.
Compared to Po$d$, each scheduler in JIQ avoids the probing operation and assigns a new task to an idle server directly as long as
its I-queue is non-empty. Due to its clear advantages, JIQ has begun to attract research attention from both industry and academia \cite{Mit16,Sto15,Sto2016,Sto2016_1}. Despite these significant research achievements made recently, 
distributed JIQ is not yet well understood from a theoretical perspective.
For example, there is no closed-form expression yet that \textit{exactly} characterizes the delay performance of distributed JIQ \cite{Mit16}\footnote{In \cite{Lu11}, the authors provided a closed-form expression that approximately characterizes the delay performance of distributed JIQ based on some simplifying assumptions. Although their expression is insightful, it is not very accurate for our system model as explained in Section~\ref{sec:simulations}.}.
Also, there seems no theoretical guarantee that the distributed JIQ (or any of its variants) is strictly better than Po$d$.

%
%shows better potential than the Po$d$ in the large-scale server system. Roughly speaking, each JIQ scheduler maintains an I-queue, where the information of some idle servers is gathered, for task scheduling. On the one hand, it avoids the server probing and reduces the communication overheads in the task scheduling. On the other hand, distributed JIQ makes tasks completed even faster under typical workloads. 
%Although it has started to atract attention from both industry and academia \cite{Mit16,Sto15,Sto2016,Sto15_1,Sto2016_1},
%%The analysis of JIQ algorithm has been studied.  Unfortunately, 
%distributed JIQ is much less understood than Po$d$. For example, there is no closed-form expression yet for distributed JIQ. 
%Additionally, there is no guarantee that distributed JIQ is strictly better than Po$d$. For instance, when the workloads become extremely heavy, JIQ performed worse than Po$d$.

\subsection{Contributions}

In this paper, we take a further step in understanding the performance of distributed JIQ. 
As our first contribution, we apply a mean-field analysis to derive semi-closed form expressions of the 
stationary tail distribution and the expected response time for distributed JIQ. 
Our expressions contain a parameter $\hat{p}_0 \in (0, 1)$ that can be efficiently calculated by a binary search.
We show that, in the large-system limit, the tail probability $\hat{s}_i$ of a server having at least $i$ tasks
is given by $\hat{s}_i = \hat{p}_0^{i-1} \lambda^i$, where $\lambda$ is the normalized arrival rate. We also show that
the expected task response time is $1 + \hat{p}_0 \sum_{i = 1}^\infty \hat{s}_i$. These two expressions allow us to
compare JIQ and Po$d$ directly and find that JIQ is not always better than Po$d$.

%Recall \cite{Mit16}, "We have not, to this point, been able to prove formally that there is unique fixed point equilibrium to these equations." 
%In our paper, we manage to obtain the unique stationary distribution of the JIQ system under the large-system limit.

As our second contribution, we propose a new variant of JIQ called JIQ-Po$d$ that strictly outperforms Po$d$. 
JIQ-Po$d$ enjoys the best of both worlds. Similar to JIQ, a scheduler with a non-empty I-queue in JIQ-Po$d$
assigns a new task to an idle server on its I-queue. Similar to Po$d$, a scheduler with an empty I-queue
in JIQ-Po$d$ probes $d$ servers and selects the least loaded one. Intuitively, JIQ-Po$d$ improves upon JIQ in that
it makes schedulers with empty I-queues ``smarter"; it improves upon Po$d$ in that schedulers with non-empty I-queues
can assign new tasks directly to idle servers without the probing operation. Using the mean-field analysis,
we are able to quantify the improvements of JIQ-Po$d$ over JIQ and Po$d$ in the large-system limit. 

\subsection{Related Work}
The Po$d$ algorithm and its variants have been widely studied and applied in today's cloud systems. 
One variant is called batch filling \cite{Lei15}, which is designed for batch arrivals. It achieves lower tail response time than the Po$d$ algorithm and guarantees a bounded maximum queue-length for the system.
Another variant is a hybrid algorithm that combines the Po$d$ with a centralized helper \cite{chunpu16}. In particular, such hybrid algorithm achieves a bounded maximum queue-length and lower response time even when the helper has a small portion of processing capacity.  
Unlike these variants, JIQ and JIQ-Po$d$ neither rely on batch arrivals nor rely on a centralized helper.

The JIQ algorithm  was originally proposed in a seminal work \cite{Lu11} in 2011. 
The authors assumed that all servers in I-queues are idle as a simplification of their performance analysis.
As pointed out in \cite{Lu11}, this assumption is violated when an idle server receives a random arrival. In this work, we do not make such assumption. Instead, we introduce delete request messages (as explained later) to ensure that all servers in I-queue are idle.

Recently, Mitzenmacher studied the distributed JIQ algorithm through a fluid-limit approach~\cite{Mit16}. He proposed an elegant classification of the states of servers and derived families of differential equations that describe the JIQ system in the large-system limit. Due to the high complexity of those differential equations, there is no expression of the equilibrium in a convenient form in terms of $\lambda$~\cite{Mit16}. 
Our work is inspired by Mitzenmacher's fluid-limit approach. By introducing delete request messages, we are able to simplify the differential equations and obtain semi-closed form expressions for distributed JIQ. Based on the insights from our analysis, 
we propose and analyzed a new variant of JIQ that outperforms both JIQ and Po$d$ in all conditions.

In a series of papers \cite{Sto15,Sto2016,Sto2016_1}, Stolyar studied a centralized JIQ algorithm where there is only one scheduler (or a fixed number of schedulers) in the system through mean-field analysis.
It shows that centralized JIQ approaches the performance of JSQ in the large-systems limit. This exciting result means that a centralized scheduler only needs to track idle servers instead of all the servers. Our work is complementary to Stolyar's work
in that we focus on distributed JIQ rather than cetralized JIQ.

Also, several recent studies explored the tradeoff between the communication overheads and the task/job response times ~\cite{revision01,revision02} as well as the tradeoff between the energy consumption and the task/job response times ~\cite{revision03} for JIQ-like algorithms. The techniques developed in this work might be useful to these studies as well.

\subsection{Organization of the Paper}

Section~\ref{sec:statement} introduces our system model and main results. In Section~\ref{sec:JIQ-analysis}, we perform the mean-field analysis for both JIQ and JIQ-Po$d$. In Section~\ref{sec:simulations}, extensive simulations are conducted to validate our analysis. Finally, Section~\ref{sec:conclusion} concludes the paper.

\section{System Model and Main Results}
\label{sec:statement}

In this section, we will introduce the system model of the distributed JIQ algorithm as well as a new variant---distributed JIQ-Po$d$ algorithm. We will then compare these two algorithms with the Po$d$ algorithm in terms of tail distribution of servers and expected task response time.

\subsection{Distributed JIQ Algorithm}
Consider a system with $N$ identical servers and $M$ schedulers, where the ratio $r$ is defined as $r = N/M$. Each scheduler is equipped with an I-queue
that stores a list of idle servers (which will be specified later).

\begin{table*}
 \caption{summary of queue length distribution under JIQ, JIQ-Po$d$ and Po$d$.}
\label{tab:compare}
\begin{tabularx}{\textwidth}{@{}l*{4}{C}c@{}}
\toprule
   &JIQ & JIQ-Po$d$   &Po$d$ \\
\addlinespace
\hline
\addlinespace

Tail distribution of server ($\hat{s}_i$ for $i \geq 1$)  

& $ \hat{p}_0^{i - 1}{\lambda ^i}$  
& ${\lambda ^{\frac{{{d^i} - 1}}{{d - 1}}}}\hat{p}_0^{\frac{{{d^{i - 1}} - 1}}{{d - 1}}}$
& $ {\lambda ^{\frac{{{d^i} - 1}}{{d - 1}}}}$ 
\\ \addlinespace \hline \addlinespace

Expected task response time ($T(\lambda)$)
   
& $1 + {\hat{p}_0}\sum\limits_{i = 1}^\infty  {{{{\hat{s}_i}}}}$    
& $1 + {\hat{p}_0}\sum\limits_{i = 1}^\infty  {{{({\hat{s}_i})}^d}}$
& $1 + \sum\limits_{i = 1}^\infty  {{{({\hat{s}_i})}^d}}$ 
\\ \addlinespace

\bottomrule
\end{tabularx}
\end{table*}

The system evolves as follows:

\begin{itemize}

\item \emph{Task arrivals}: Tasks arrive at the system according to a Poisson process of rate $\lambda N$, where $\lambda < 1$, and are sent to a scheduler uniformly at random. Thus, each scheduler observes a Poisson arrival process of rate $\lambda N/M$.

\item \emph{Schedulers with I-queues}: Each scheduler has an I-queue, which maintains a list of idle servers.  Upon a task arrival, each scheduler checks its I-queue and assigns 
the task to a server according to the following rule: 
If the I-queue is non-empty, the scheduler selects an idle server uniformly at random from its I-queue. If the I-queue is empty, the scheduler 
selects a server uniformly at random from all the servers. 

\item \emph{Servers}: Each server has an infinite buffer and processes tasks in a first-in first-out (FIFO) manner. The task processing times are exponentially distributed with mean~$1$. Whenever a server becomes idle, it joins an I-queue selected uniformly at random among all I-queues.
Whenever an idle server becomes busy, it leaves its associated I-queue in one of the following two ways:
	  \begin{figure}[tp]
	    \centering
	  \includegraphics[width = 0.48\textwidth]{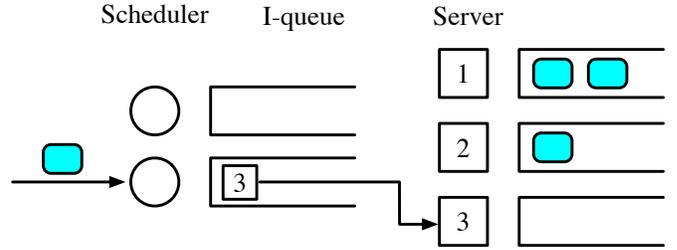}
	  %\vspace{-3mm}
	  \caption{Server $3$ is selected by Scheduler $2$ and leaves its I-queue, where $N=3$ and $M=2$.}
	  \label{fig:JIQ}
	  %\vspace{-4mm}
	  \end{figure}
	
	\begin{figure}[tp]
	   \centering	
	\includegraphics[width = 0.48\textwidth]{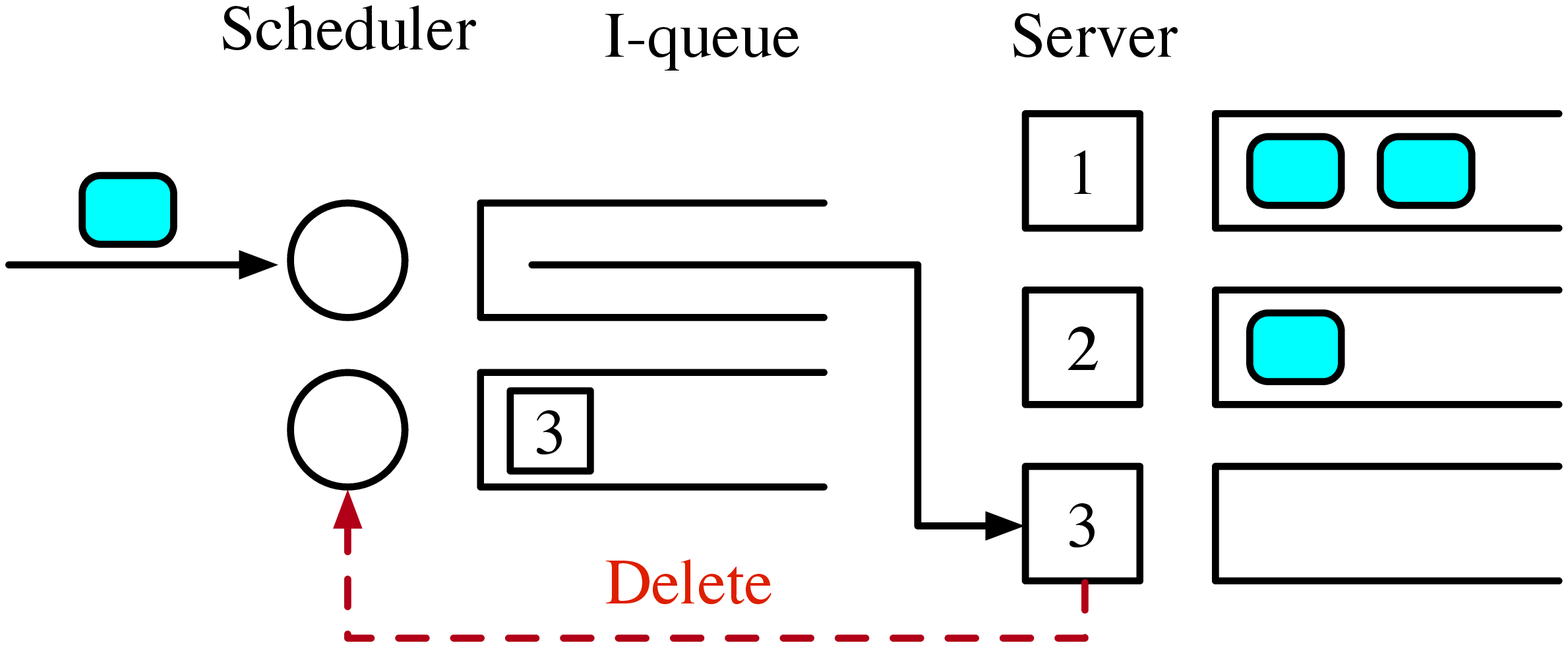}
	%\vspace{-3mm}
	\caption{Server $3$ is selected by Scheduler $1$ and sends a ``delete request'' message to Scheduler $2$, where $N=3$ and $M=2$.}
	\label{fig:JIQ-1}
	%\vspace{-4mm}
	\end{figure}
	
\begin{enumerate}
\item If it is selected by a scheduler with a non-empty I-queue, then it simply leaves the I-queue, as shown in Figure~\ref{fig:JIQ}.

\item If it is selected by a scheduler with an empty I-queue, then it has to inform its associated I-queue by sending a  ``delete request'' message, as shown in Figure~\ref{fig:JIQ-1}\footnote{We observe that the same strategy is used in \cite{Sto2016}, where the idle server informs its associated I-queue to destroy its identity in the list.}.

\end{enumerate}
% \textcolor{red}{Hence, a server joins in just one I-queue if and only if it is idle.}
\end{itemize}

\begin{remark}
We note that some distributed JIQ algorithm doesn't use the ``delete request'' messages (e.g., in \cite{Mit16}), allowing I-queues having non-idle servers.
Although it reduces the communication overhead, it complicates the theoretical analysis. As we will see in Section~\ref{sec:impact}, such extra communication overhead is acceptable. 
\end{remark}

\subsection{Distributed JIQ-Po$d$ Algorithm}

The distributed JIQ algorithm described above doesn't always outperform the Po$d$ algorithm, especially under heavy workload where most I-queues are empty. 
% To address this issue, we propose a new variant of JIQ, namely JIQ-Po$d$, which 
% combines the advantages of JIQ and Po$d$. It works as follows. 
To address this issue, we propose a new variant of JIQ, namely JIQ-Po$d$, which combines the advantages of JIQ and Po$d$. It works as follows.

\begin{figure}[htp]
   \centering

\includegraphics[width = 0.48\textwidth]{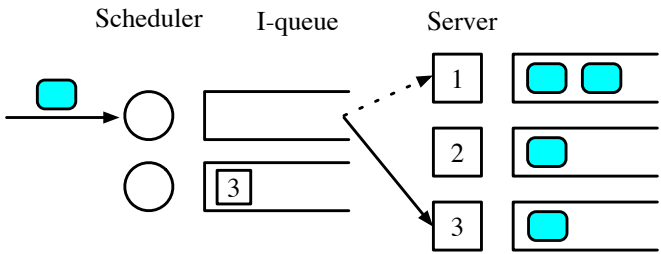}
%\vspace{-3mm}
\caption{Scheduler $1$ probes Server $1$ and Server $3$, and assigns a new task to Server $3$, where $N=3$ and $M=2$.}
\label{fig:JIQ-2}
%\vspace{-4mm}
\end{figure}

{\bf Schedulers under JIQ-Po$d$.} Upon a task arrival, each scheduler checks its I-queue and assigns 
the task to a server according to the following rule: 
If its I-queue is non-empty, the scheduler selects an idle server uniformly at random from its I-queue. If its I-queue is empty,  the scheduler probes $d$ servers uniformly at random and assigns the task to the least loaded one, as shown in Figure~\ref{fig:JIQ-2}. Essentially, each scheduler with an empty I-queue is applying the Po$d$ strategy.

% Whenever a new task is sent to a scheduler with an empty I-queue, the scheduler probes $d$ servers uniformly at random and assigns the task to the least loaded one, as shown in Figure~\ref{fig:JIQ-2}. That is, each scheduler with an empty I-queue is applying the Po$d$ strategy. 
All other steps remain the same  as the distributed JIQ algorithm.
Clearly, when $d=1$, our JIQ-Po$d$ algorithm reduces to the distributed JIQ.

\subsection{Main Results}

Table~\ref{tab:compare} presents our main results that characterize the stationary tail distribution and the expected task response time
in the large-system limit (i.e., $N \rightarrow \infty$ and $M \rightarrow \infty$ with the ratio $r=N/M$ fixed), where
$\hat{p}_0$ is some parameter in $(0,1)$ (which will be specified later).
The stationary tail distribution $\hat{s}_i$ is the fraction of servers having no less than $i$ tasks in their task queues. 
(Note that $\hat{s}_0$ is always equal to $1$, and that the $\{ \hat{s}_i \}$ are non-increasing.) 
The smaller $\hat{s}_i$ is, the shorter the task delay.
The expected task response time $T(\lambda)$ measures the average completion time for a task in steady state. 
%Note that we deem $p_0$ as a parameter, which is in $(0,1)$. %According to Figure, $p_0$ is no moe than $0.5$ when arrival rate $0.95$. 
%As shown in Table~\ref{tab:compare} and Figure~\ref{fig:main_reslut}, we conclude that 

First, we observe that JIQ-Po$d$ gives the best tail distribution $\hat{s}_i$. 
Compared to Po$d$, JIQ-Po$d$ has an additional factor of ${\hat{p}_0^{\frac{{{d^{i - 1}} - 1}}{{d - 1}}}} < 1$, since $\hat{p}_0
\in (0, 1)$. For instance, when $d=2$, $i=2$ and $p_0 = 0.5$, such factor equals to $0.5$. 
Compared to JIQ, JIQ-Po$d$ has larger exponents of $\lambda$ and $\hat{p}_0$. For example, when $d=2$ and $i=3$, the exponent of $\lambda$ under JIQ-Po$d$ is ${\frac{{{d^i} - 1}}{{d - 1}}} = 7 > i = 3$, and the exponent of $\hat{p}_0$ under JIQ-Po$d$ is ${\frac{{{d^{i - 1}} - 1}}{{d - 1}}} = 3 > i-1 = 2$.  
\begin{figure}[htp]
   \centering

\includegraphics[width = 0.50\textwidth]{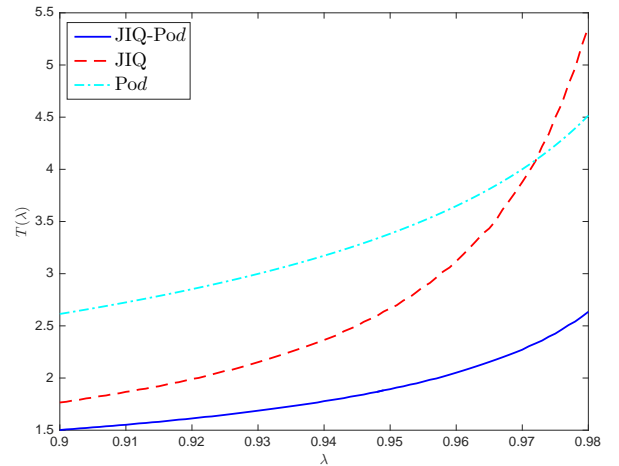}
%\vspace{-3mm}
\caption{Comparison of the expected task response time among JIQ, JIQ-Po$d$ and Po$d$, when $r = 10$ and $d=2$.}
\label{fig:main_reslut}
%\vspace{-4mm}
\end{figure}

Second, we observe that JIQ-Po$d$ achieves the shortest expected task response time $T(\lambda)$. 
Compared to Po$d$, JIQ-Po$d$ has an additional factor of $\hat{p}_0 <1$.
Compared to JIQ, JIQ-Po$d$ has larger exponents of $\hat{s}_i$. 
To better illustrate the advantage of JIQ-Po$d$, we provide a concrete numerical example in Figure~\ref{fig:main_reslut},
which shows that, when $\lambda = 0.98$, $T(\lambda)$ of JIQ-Po$d$ is only around $2.6$, whereas $T(\lambda)$ of JIQ and Po$d$ are $5.3$ and $4.5$, respectively.

\section{Mean-Field Analysis}
\label{sec:JIQ-analysis}
In this section, we will use the mean-field analysis to study
the stationary distributions of the queue lengths,
as well as the corresponding expected task response time, under JIQ. We will then apply the same analysis to JIQ-Po$d$.
The underlying assumptions behind the mean-field analysis will be validated through simulations
in Sec.~\ref{sec:verification}. In fact, these assumptions can be rigorously validated using proof techniques
such as Kurtz's theorem, which is beyond the scope of this paper\footnote{The evolution of the system state
can be characterized by a density-dependent continuous-time Markov chain ${\widetilde {\mathbf{Q}}^{(N)}}(t)$. 
This allows us to apply Kurtz's theorem to rigorously validate the
use of the mean-field analysis. Most of the proof steps towards this direction are standard except for the step showing
the global convergence of the underlying ordinary differential equations.}.

First, we look at the state of a single server in the system. Let $\left(X_i^{(N)}(t), Y_i^{(N)}(t) \right)$ denote the \emph{state} of the $i$th server at time $t$ in a system of $N$ servers
and $M$ I-queues, where $X_i^{(N)}(t)$ is the queue length of the $i$th server at time $t$ and
$Y_i^{(N)}(t)$ is the index of the associated I-queue. If the $i$th server doesn't belong to any I-queue at time $t$, 
we set $Y_i^{(N)}(t) = 0$. It is easy to check that $\left\{ \left(X_i^{(N)}(t), Y_i^{(N)}(t) \right)\right \}_{i=1}^N$ forms 
an irreducible, aperiodic, continuous-time Markov chain
under our system model. Moreover, the following theorem shows that 
this Markov chain is positive recurrent.
%and thus has a unique stationary distribution.
 Thus, it has a unique stationary distribution.

\begin{theorem}\label{thm:positive-recurrent}
 The Markov Chain $\left\{ \left(X_i^{(N)}(t), Y_i^{(N)}(t) \right)\right \}_{i=1}^N$ under the JIQ algorithm is positive recurrent. 
\end{theorem}

\begin{IEEEproof}
We first construct a potential function and then apply Foster-Lyapunov theorem. Please see Appendix~\ref{apx-a} for details. 
\end{IEEEproof}

% In order to conduct the mean-field analysis, we need to introduce a new representation of the system state.
We now introduce a new representation of the system state to conduct the mean-field analysis.
Let $Q_i^{(N)}(t)$ denote the number of servers with $i$ tasks at time $t$.
Let $Q_{(0,j)}^{(N)}(t)$ denote the number of idle servers that belong to I-queues of size $j$ at time $t$. Then, the system state at time $t$
can be described by
\[{{\mathbf{Q}}^{(N)}}(t) = \left\{ {Q_{(0,1)}^{(N)}(t),Q_{(0,2)}^{(N)}(t), \cdots ,Q_1^{(N)}(t),Q_2^{(N)}(t), \cdots } \right\}.\]
One can verify that ${{\mathbf{Q}}^{(N)}}(t)$ also forms a continuous-time Markov chain under our system model, because
the individual servers (or I-queues) of the same queue-length are \emph{indistinguishable} for system evolution.
In other words, our new Markov chain ${{\mathbf{Q}}^{(N)}}(t)$ captures the ``essential" information of our original Markov chain 
$\left\{ \left(X_i^{(N)}(t), Y_i^{(N)}(t) \right)\right \}_{i=1}^N$. In particular, if our original Markov chain has a unique stationary
distribution, so does our new Markov chain.

For convenience, we further introduce a normalized version of ${{\mathbf{Q}}^{(N)}}(t)$ defined as 
\[
{\widetilde {\mathbf{Q}}^{(N)}}(t) = \frac{1}{N} {{\mathbf{Q}}^{(N)}}(t).
\]
Note that $\tilde Q_{(0, j)}^{(N)}(t)$ is the fraction of servers that belong to I-queues of size $j$ at time $t$,
and $\tilde Q_i^{(N)}(t)$ is the fraction of servers with $i$ tasks at time $t$.
Clearly, ${\widetilde {\mathbf{Q}}^{(N)}}(t)$ is also positive recurrent and has a unique stationary distribution.
In addition, we can show that ${\widetilde {\mathbf{Q}}^{(N)}}(t)$ is density dependent.

The mean-field analysis proceeds as follows. We assume that the $N$ servers are in the steady state.
We also assume that the states of these servers are identically and independently distributed (i.i.d.) \cite{Lei15}.
This i.i.d. assumption will be validated later through simulations. We now consider the state evolution of one server in the system
under the i.i.d. assumption. Note that the possible server states are from the set
\[
\{ (0, 1), (0, 2), \ldots, 1, 2, \ldots \}
\]
where the state-$(0, j)$ means that the server is idle and belongs to an I-queue of size $j$, 
and the state-$i$ means that the server has $i$ tasks in its queue. Let
\[{\mathbf{q}} = \left\{ {{q_{(0,1)}},{q_{(0,2)}}, \cdots ,{q_1},{q_2}, \cdots } \right\}\]
denote the stationary distribution of the server state. (Note that ${\mathbf{q}}$ is unique because
${\widetilde {\mathbf{Q}}^{(N)}}(t)$ is positive recurrent.) Then, by the Strong Law of Large Numbers,
$q_{0,j}$ can be interpreted as the fraction of servers belonging to an I-queue of size $j$,
and $q_i$ can be interpreted as the fraction of servers with $i$ tasks in the large-system limit.
This means that the stationary distribution of ${\widetilde {\mathbf{Q}}^{(N)}}(t)$ ``concentrates" on
${\mathbf{q}}$ as $N \to \infty$. 

We are now ready to derive the stationary distributions under JIQ and JIQ-Po$d$ in the large-system limit.

%Next, we characterize the set of state transitions for the Markov chain ${{\bf{Q}}^{(N)}}(t)$ and its normalized version
%${\widetilde {\bf{Q}}^{(N)}}(t)$, which will be used later in the mean-field analysis.

\subsection{The Stationary Distribution Under JIQ}
In this subsection, we will derive the stationary distribution of one server in the system under JIQ.
The i.i.d. assumption described above allows us to obtain the transition rates for the state evolution of the server.
In fact, this assumption holds asymptotically in the large-system limit. In other words, the stationary distribution 
derived here is sufficiently accurate for large-scale systems.

To derive the transition rates, we need some additional notations.
Let $p_i$ be the fraction of I-queues of size $i$ in the large-system limit. Then we have

%large-system limit (i.e., $N \rightarrow \infty$ and $M \rightarrow \infty$ with the ratio $r=M/N$ fixed).
%In the large-system limit, let ${\mathbf{q}}$ be the stationary distribution of a single server.
%\[{\mathbf{q}} = \left\{ {{q_{(0,1)}},{q_{(0,2)}}, \cdots ,{q_1},{q_2}, \cdots } \right\},\]
%where ${q_{(0,i)}}$ denotes the probability that it has $0$ tasks and joins an I-queue with queue-length $i$; ${q_i}$ denotes the probability that its queue-length equals to $i$. 
%Each server forms an independent Markov chain, as shown in Figure~\ref{fig:JIQ-transition}.
%
%\begin{figure}[tp]
%   \centering
%
%\includegraphics[width = 0.45\textwidth]{transition-rate.png}
%%\vspace{-3mm}
%\caption{The Markov chain of a single server under the JIQ algorithm.}
%\label{fig:JIQ-transition}
%%\vspace{-4mm}
%\end{figure}

\[
p_i = \left\{ \begin{array}{ll}
\frac{{r{q_{(0,i)}}}}{i}, & i \ge 1,\\
1 - \sum\limits_{j = 1}^\infty  \frac{{r{q_{(0,j)}}}}{j}, & i = 0
\end{array} \right.
\]
where the first equation follows from the fact that the number of servers in state-$(0, i)$
is equal to $i$ times the number of I-queues of size $i$, and the second equation follows from the fact that
$\sum_{j = 0}^\infty p_j = 1$.

We now derive the transition rates for the state evolution of a single server as follows:
%\[\left\{ \begin{array}{l}
%{r_{i,i - 1}} = 1,\\
%{r_{i - 1,i}} = \lambda {p_0},\\
%{r_{1,(0,j)}} = {p_{j - 1}},\\
%{r_{(0,j),1}} = \lambda ({p_0} + \frac{r}{j}),\\
%{r_{(0,j - 1),(0,j)}} = r{q_1},\\
%{r_{(0,j),(0,j - 1)}} = \lambda (j - 1)({p_0} + \frac{r}{j}),
%\end{array} \right.\]
%where $i \geq 2$ and $j \geq 1$.

\begin{itemize}
\item ${r_{i,i - 1}} = 1$ for $i \ge 2$. \\
	The processing rate of a task is exponentially distributed with mean $1$.
\item ${r_{i - 1,i}} = \lambda {p_0}$ for $i \ge 2$.\\ 
	The task arrival rate is $\lambda N$, the probability of joining an empty I-queue is $p_0$, and the probability of selecting the target server over all servers is $\frac{1}{N}$.
\item ${r_{1,(0,j)}} = {p_{j - 1}}$ for $j \ge 1$. \\
	The processing rate of a task is exponentially distributed with mean $1$, and the probability of joining an I-queue of size $j-1$ is $p_{j-1}$.
\item ${r_{(0,j),1}} = \lambda ({p_0} + \frac{r}{j})$ for $j \ge 1$.\\ 
	The task arrival rate is $\lambda N$. There are two events leading to this state change because of a newly arrival task. 
	The first event is that the new task is routed to an empty I-queue and then directed to the target server. 
	The probability of this event is $p_0 \cdot \frac{1}{N}$. The second event is that the new task is routed to the I-queue associated
	with the target server
	 and then directed to the target server. The probability of this event is
	$\frac{1}{M} \cdot \frac{1}{j} = \frac{r}{N} \cdot \frac{1}{j}$. Hence, the transition rate is $\lambda N \left( p_0 \cdot \frac{1}{N} + 
	  \frac{r}{N} \cdot \frac{1}{j} \right) $.
\item ${r_{(0,j - 1),(0,j)}} = r{q_1}$ for $j \ge 2$. \\
	The generating rate of idle servers is $q_1 N$, and the probability of selecting the I-queue associated with the target 
	server is $\frac{1}{M}$.
\item ${r_{(0,j),(0,j - 1)}} = \lambda (j - 1)({p_0} + \frac{r}{j})$ for $j \ge 2$.\\ 
	The task arrival rate is $\lambda N$. There are two events resulting in this state change.
	The first event is that the new task is routed to an empty I-queue and then directed to an idle server having the same I-queue
	as the target server. The probability of this event is $p_0 \cdot \frac{j-1}{N}$.
	The second event is that the new task is routed to the I-queue associated with the target server and then directed to another idle server. The probability of this event is $\frac{1}{M} \cdot \frac{j-1}{j}$. Hence, the transition rate is
	$\lambda N \left(  p_0 \cdot \frac{j-1}{N} + \frac{r}{N} \cdot \frac{j-1}{j} \right)$.
	 \end{itemize}

Based on the above transition rates, one can easily write down the local balance equations as
\begin{equation}\label{server_side_calculate}
\left\{ \begin{array}{l}
{q_i}{r_{i,i - 1}} = {q_{i - 1}}{r_{i - 1,i}}, \mbox{ for } i \ge 2,\\
{q_{(0,j)}}{r_{(0,j),1}} = {q_1}{r_{1,(0,j)}},\mbox{ for } j \ge 1,\\
{q_{(0,j)}}{r_{(0,j),(0,j - 1)}} = {q_{(0,j - 1)}}{r_{(0,j - 1),(0,j)}},\mbox{ for } j \ge 2.
\end{array} \right.
\end{equation}

The following theorem computes the stationary distribution of the state of a single server
in the large-system limit by finding a particular distribution that satisfies the local balance equations \eqref{server_side_calculate}.

\begin{theorem} \label{thm:main}
The stationary distribution of the state of a single server under JIQ in the large-system limit is
\begin{equation}\label{JIQ_steady}
	\left\{ \begin{array}{l}
{\hat{q}_{(0,i)}} = \frac{{{r^{i - 1}}{{(1 - \lambda {\hat{p}_0})}^i}}}{{\mathop \prod \limits_{j = 1}^i (r + j{\hat{p}_0})}}i{\hat{p}_0},{\mbox{ for }}i \ge 1,\\
{\hat{q}_i} = \hat{p}_0^{i - 1}{\lambda ^i}(1 - {\hat{p}_0}\lambda ),{\mbox{ for }}i \ge 1
\end{array} \right.
\end{equation}
where $\hat{p}_0$ is the unique solution to the following equation 
\begin{equation}\label{eq:p0}
%p_0 + \sum\limits_{i = 1}^\infty  {\frac{{r{q_{(0,j)}}}}{j}}  = 1
p_0 + \sum\limits_{i = 1}^\infty \frac{{{r^{i}}{{(1 - \lambda {{p}_0})}^i}}}{{\mathop \prod \limits_{j = 1}^i (r + j{{p}_0})}}{{p}_0} = 1
\end{equation}
over the interval $(0, 1)$.
\end{theorem}

\begin{remark}
Let 
\[f(p_0) \triangleq p_0 + \sum\limits_{i = 1}^\infty \frac{{{r^{i}}{{(1 - \lambda {{p}_0})}^i}}}{{\mathop \prod \limits_{j = 1}^i (r + j{{p}_0})}}{{p}_0}.
\]
Then \eqref{eq:p0} can be written as $f(p_0) = 1$.
Interestingly, $f(p_0)$ can be expressed in terms of Gamma functions 
\begin{multline}\label{eq:gamma}
  f({p_0}) = {p_0} + {e^{r( - \lambda  + \frac{1}{{{p_0}}})}}{\left[r\left( - \lambda  + \frac{1}{{{p_0}}}\right)\right]^{ - \frac{r}{{{p_0}}}}}\\
 \times \left[ {\Gamma\left(\frac{{r + {p_0}}}{{{p_0}}}\right) - \Gamma\left(\frac{{r + {p_0}}}{{{p_0}}},r\left( - \lambda  + \frac{1}{{{p_0}}}\right)\right)} \right]  p_0
\end{multline}
where the Gamma functions $\Gamma(x)$ and $\Gamma(x,a)$ are respectively defined as
\[\Gamma(x) = \int_0^\infty  {{t^{x - 1}}{e^{ - t}}dt} \]
and
\[\Gamma(x,a) = \int_a^\infty  {{t^{x - 1}}{e^{ - t}}dt}. \]
To see this, notice that
\begin{align*}
\Gamma(x) - \Gamma(x,a) &= \int_0^a  {{t^{x - 1}}{e^{ - t}}dt}   \\
&= a^x e^{- a} \sum_{k = 0}^{\infty} \frac{a^k}{x (x+1) \cdots (x + k)}.
\end{align*}
Setting $x = \frac{{r + {p_0}}}{p_0}$ and $a = r\left( - \lambda  + \frac{1}{{{p_0}}}\right)$ gives the expression
of $f(p_0)$.
\end{remark}

\begin{IEEEproof}
We will prove Theorem~\ref{thm:main} through two steps. First, we will show that ~\eqref{eq:p0} indeed has a unique solution.
Second, we will show that the distribution $\mathbf{\hat{q}}$ satisfies the local balance equations~\eqref{server_side_calculate}.

To prove the first step, we will show in Appendix~\ref{apx-b} that $f(p_0)$ is differentiable and monotonically increasing over the interval $(0, 1)$.
Notice that $f(0) = 0$ and $f(1) > 1$ when $\lambda < 1$. Hence, by the Intermediate Value Theorem, the equation $f(p_0) = 1$ has a unique solution over the interval $(0, 1)$.

To prove the second step, we only need to verify that the distribution $\mathbf{\hat{q}}$ (constructed in \eqref{JIQ_steady}) satisfies the local balance  equations~\eqref{server_side_calculate}. This verification is straightforward.
\end{IEEEproof}

\begin{remark}
In order to numerically compute the value of $\hat{p}_0$, we consider
a truncated version of $f(p_0)$ defined as
\[
f_n(p_0) \triangleq p_0 + \sum\limits_{i = 1}^n \frac{{{r^{i}}{{(1 - \lambda {{p}_0})}^i}}}{{\mathop \prod \limits_{j = 1}^i (r + j{{p}_0})}}{{p}_0}.
\] 
Intuitively, $f_n(p_0)$ tends to $f(p_0)$ as $n$ increases, because the terms of $f(p_0)$ are decreasing exponentially to $0$. 
We can bound the ``approximation error" $f(p_0) - f_n(p_0)$ as follows:
\begin{align*}
f(p_0) - f_n(p_0) &= \sum\limits_{i = n+1}^\infty  {\frac{{{r^i}{{(1 - \lambda {p_0})}^i}}}{{\prod\limits_{j = 1}^i {(r + j{p_0})} }}} {p_0} \\
&= {\frac{{{r^n}{{(1 - \lambda {p_0})}^n}}}{{\prod\limits_{j = 1}^n {(r + j{p_0})} }}} {p_0} \cdot
\sum_{i = 1}^\infty \frac{r^i (1 - \lambda {p_0})^i}{\prod_{j = n + 1}^{n + i} (r + j{p_0})},
\end{align*}
where ${\frac{{{r^n}{{(1 - \lambda {p_0})}^n}}}{{\prod_{j = 1}^n {(r + j{p_0})} }}} {p_0}$ is the last term
of $f_n(p_0)$. Note that
\[
    \sum_{i = 1}^\infty \frac{r^i (1 - \lambda {p_0})^i}{\prod_{j = n + 1}^{n + i} (r + j{p_0})} < 
\sum_{i = 1}^\infty \frac{r^i (1 - \lambda {p_0})^i}{(r + n{p_0})^i} 
= \frac{r(1 - \lambda {p_0})}{(n + r \lambda) p_0}.
\]
 Hence, the approximation error is less than a fraction of $\frac{r(1 - \lambda {p_0})}{(n + r \lambda) p_0}$ of
the last term of $f_n(p_0)$, which is negligible for large $n$. In fact, our numerical simulation suggests that $f_n(p_0)$ is sufficiently close to $f(p_0)$ and is monotonically increasing when $n > 20$. This allows us to 
apply a simple binary search to solve the equation $f_n(p_0) = 1$.
\end{remark}

We can derive the stationary tail  distribution and the expected task response time based on Theorem~\ref{thm:main}.
\begin{corollary}\label{cor:mean-task-response}
    In the large-system limit, the stationary tail  distribution under JIQ is
\[{\hat{s}_i} = \left\{ \begin{array}{l}1, \mbox{ for } i = 0,\\\hat p_0^{i - 1}{\lambda ^i}, \mbox{ for } i \ge 1\end{array} \right.\]
      and the expected task response time under JIQ is $\frac{1}{{1 - {\hat{p}_0}\lambda }}$.
\end{corollary}

\begin{IEEEproof}
The stationary tail distribution 
\[
{\hat{s}_i} = \sum\limits_{j = i}^\infty  {{\hat{q}_j}} = 
	\sum\limits_{j = i}^\infty  {\hat p_0^{j - 1}{\lambda ^j}(1 - {{\hat p}_0}\lambda )}  = \hat{p}_0^{i - 1}{\lambda ^i}.
\]
This proves the first part. To compute the expected task response time, we consider the following two cases.
\begin{enumerate}
\item A new task is sent to a non-empty I-queue (with probability $1-\hat{p}_0$). The expected task response time in this case is $1$.
\item A new task is sent to an empty I-queue (with probability $\hat{p}_0$). The 
expected task response time in this case is $\sum\limits_{i = 0}^\infty  {(i + 1){\hat{q}_i}}$.
\end{enumerate}
    
    Combining the above two cases, the expected task response time is
    \[(1 - {\hat{p}_0}) + {\hat{p}_0}\sum\limits_{i = 0}^\infty  {(i + 1){\hat{q}_i}}  = \frac{1}{{1 - {\hat{p}_0}\lambda }}.\]
    This completes the second part.
\end{IEEEproof}

\subsection{The Stationary Distribution Under JIQ-Po$d$}

%Similar as that of our JIQ algorithm, the $\left\{ \left(X_i^{(N)}(t), Y_i^{(N)}(t) \right)\right \}_{i=1}^N$ forms 
%an irreducible, aperiodic, continuous-time Markov chain
%under JIQ-Po$d$. Moreover, the following theorem shows that 
%this Markov chain is positive recurrent
%and thus has a unique stationary distribution.
%
%\begin{theorem}\label{thm:positive-recurrent-JIQ-extension}
% The Markov Chain $\left\{ \left(X_i^{(N)}(t), Y_i^{(N)}(t) \right)\right \}_{i=1}^N$ under the JIQ-Po$d$ algorithm is positive recurrent. 
%\end{theorem}
%
%%\begin{theorem}\label{thm:positive-recurrent-JIQ-extension}
%% The Markov Chain ${ \{ {\mathbf{Q}^{(N)}}(t), {\mathbf{P}^{(M)}}(t)\} _{t \geq 0}}$ is positive recurrent.
%%\end{theorem}
%
%The proof of Theorem~\ref{thm:positive-recurrent-JIQ-extension} follows the standard Foster-Lyapunov theorem. 
%%Hence, ${ \{ {\mathbf{Q}^{(N)}}(t), {\mathbf{P}^{(M)}}(t)\} _{t \geq 0}}$ has a unique stationary distribution.
%
%Recall the i.i.d. assumption of a individual server in the large-system limit, we then focus on the transitions of a single server. The corresponding transition rates go as follows. 
Similar to our previous analysis for JIQ, we can derive the transition rates for JIQ-Po$d$ as follows:
%We derive the transition rates for the state evolution of a single server as follows:
\[\left\{ \begin{array}{l}
{r_{i,i - 1}} = 1,\mbox{ for } i \ge 2,\\
\lambda {p_0}\left[ {\frac{{{{\left( {\sum\limits_{j = i - 1}^\infty  {{q_j}} } \right)}^d} - {{\left( {\sum\limits_{j = i}^\infty  {{q_j}} } \right)}^d}}}{{{q_{i - 1}}}}} \right] ,\mbox{ for } i \ge 2,\\
{r_{1,(0,j)}} = {p_{j - 1}},\mbox{ for } j \ge 1,\\
{r_{(0,j),1}} = \lambda \left[ {{p_0}\frac{{1 - {{\left( {\sum\limits_{j = i - 1}^\infty  {{q_j}} } \right)}^d}}}{{{q_0}}} + \frac{r}{j}} \right],\mbox{ for } j \ge 1,\\
{r_{(0,j - 1),(0,j)}} = r{q_1},\mbox{ for } j \ge 2.\\
{r_{(0,j),(0,j - 1)}} = \lambda (j - 1)\left[ {{p_0}\frac{{1 - {{\left( {\sum\limits_{j = i - 1}^\infty  {{q_j}} } \right)}^d}}}{{{q_0}}} + \frac{r}{j}} \right],\mbox{ for } j \ge 2.\\
\end{array} \right.\]

%Following a similar analysis as that of the JIQ algorithm, the following theorem states the stationary distribution under the JIQ-Po$d$ algorithm in the large-system limit.
The local balance equations are the same as those in ~\eqref{server_side_calculate}.
Based on ~\eqref{server_side_calculate}, we can calculate the stationary distribution of the status of single server in the large-system limit.

\begin{theorem} \label{thm:main-extension}
The stationary distribution of the state of a single server under JIQ-Po$d$ in the large-system limit is
\begin{equation}\label{JIQ_steady2}
	\left\{ \begin{array}{l}
{\hat{q}_{(0,i)}}=\frac{{{r^{i - 1}}{{(1 - {\lambda ^d}{\hat{p}_0})}^i}}}{{\mathop \prod \limits_{j = 1}^i \left[ {r + j{\hat{p}_0}(\frac{{1 - {\lambda ^d}}}{{1 - \lambda }})} \right]}}i{\hat{p}_0},{\mbox{ for }}i \ge 1,\\
{\hat{q}_i} = {\lambda ^{\frac{{{d^i} - 1}}{{d - 1}}}}{\hat{p}_0^{\frac{{{d^{i - 1}} - 1}}{{d - 1}}}}(1 - {\lambda ^{{d^i}}}{\hat{p}_0^{{d^{i - 1}}}}),
   \mbox{ for } i \ge 1
\end{array} \right.
\end{equation}
where $\hat{p}_0$ is the unique solution to the following equation 
\begin{equation}\label{eq:p01}
%p_0 + \sum\limits_{i = 1}^\infty  {\frac{{r{q_{(0,j)}}}}{j}}  = 1
p_0 + \sum\limits_{i = 1}^\infty \frac{{{r^{i}}{{(1 - \lambda {{p}_0})}^i}}}{{\mathop \prod \limits_{j = 1}^i (r + j{{p}_0})}}{{p}_0} = 1
\end{equation}
over the interval $(0, 1)$.
\end{theorem}

\begin{IEEEproof}
The proof is omitted here  as it is essentially the same as the proof for Theorem~\ref{thm:main}.
% due to the space constraints.	
\end{IEEEproof}

%Basically, the proof of Theorem~\ref{thm:main-extension} follows the two steps of Theorem~\ref{thm:main}. % The main difference of the stationary distribution construction under the JIQ-Po$d$ results from the different local balance equations. 
%Due to the page limit, the rest of the proof is omitted here.

%According to the above transition rates, we have the following global balance equations, where the $q_{(0,i)}$ is represented through $p_i$:
%
%\begin{equation}\label{server_side_calculate_extension}
%\left\{ \begin{gathered}
%  \begin{split} &\lambda {p_0}\left[ {{{(\sum\limits_{j = i - 1}^\infty  {{q_j}} )}^d} - 2{{(\sum\limits_{j = i}^\infty  {{q_j}} )}^d} + {{(\sum\limits_{j = i + 1}^\infty  {{q_j}} )}^d}} \right] \\ &= ({p_{i - 1}} - {p_i}), \mbox{ for } i \geq 2 ,\end{split} \hfill \\
%    \begin{split} &\lambda {p_0}\left[ {{{(\sum\limits_{j = 0}^\infty  {{q_j}} )}^d} - 2{{(\sum\limits_{j = 1}^\infty  {{q_j}} )}^d} + {{(\sum\limits_{j = 2}^\infty  {{q_j}} )}^d}} \right] + \lambda (1 - {p_0}) \\ &=({p_{i - 1}} - {p_i}), \end{split}\hfill \\
%	  \begin{split} &{ - \lambda {p_0}\left[ {\frac{{1 - {{(1 - {q_0})}^d}}}{{1 - (1 - {q_0})}}} \right](i{p_i} - (i + 1){p_{i + 1}})} +\lambda r({p_i} - {p_{i + 1}}) \\ &= r{q_1}({p_{i - 1}} - {p_i}), \mbox{ for } i \geq 1, \end{split} \hfill \\
%	  \lambda r{p_1} + \lambda {p_0}\left[ {\frac{{1 - {{(1 - {q_0})}^d}}}{{1 - (1 - {q_0})}}} \right]{p_1} = r{p_0}{q_1}. \hfill \\
%\end{gathered}  \right.
%\end{equation}
%Also, we have the same boundary conditions as Equation~\eqref{boundary-condition}. 

\begin{corollary}

    In the large-system limit, the stationary tail distribution under JIQ-Po$d$ is
\[{\hat{s}_i} = \left\{ \begin{array}{l}1, \mbox{ for } i = 0,\\{\lambda ^{\frac{{{d^i} - 1}}{{d - 1}}}}\hat{p}_0^{\frac{{{d^{i - 1}} - 1}}{{d - 1}}}, \mbox{ for } i \ge 1\end{array} \right.\]
    and the expected task response time under the JIQ-Po$d$ algorithm is $1 + {\hat{p}_0}\sum\limits_{i = 0}^\infty  {{{({\hat{s}_i})}^d}}$.
\end{corollary}
\begin{IEEEproof}
The proof is omitted here  as it is essentially the same as the proof for Corollary ~\ref{cor:mean-task-response}.
% due to the space constraints.	
\end{IEEEproof}

Under the JIQ-Po$d$ algorithm, the tail distribution of server queue length is lighter. Hence, task is processed faster.

\section{simulations}
\label{sec:simulations}

 In this section, we will validate our theoretical results, measure the impact of ``delete request'' messages and compare various JIQ algorithms in  finite-sized systems. In all of our simulations, we start from an empty system with the number of servers, $N$, set to be either $500$ or $1000$. 
 The number of I-queues, $M$, is chosen as $M=\frac{N}{r}$, where $r$ will be specified later. 
 The simulation results are based on the average of $10$ runs, where each run lasts for $100,000$ unit times.
 
 % the performance of the JIQ algorithm and the JIQ-Po$d$ algorithm by simulations. In particular, we first validate our theoretical results through simulations with limited size systems. Then, we compare the performance of JIQ-Po$d$ with various other JIQ extensions. We set up the server system with two different system sizes: 500 servers and 1000 servers. The I-queue size of the system varies with certain ratio $r$. The processing rate of a single server is $1$ unit time. When the simulation starts, the whole server system is empty. Then, it runs for $10000$ units of time.

\subsection{Validation of the Mean-Field Analysis}\label{sec:verification}
In this subsection, we evaluate our predictions for the stationary distribution and the expected task response time.

%First, we focus on the stationary distribution of states,  ${\mathbf{q}}$. 
%Recall that ${\mathbf{q}}$ can represent the fraction of servers of different states in the large-system limit,
 Table~\ref{tab:verify} compares the stationary distributions for JIQ and JIQ-Po$d$ obtained from prediction and simulation.
We observe that the larger the system size is, the higher the accuracy becomes. 
When the server size is only $500$, the maximum relative error rate under JIQ is as small as $3.3\%$ for $\hat{q}_5$.

\begin{table*}
\newcommand{\tabincell}[2]{\begin{tabular}{@{}#1@{}}#2\end{tabular}}
 \centering
 \caption{Prediction versus Simulations for JIQ and JIQ-Po$d$ when $r= 10$ and $\lambda = 0.9$.}
\label{tab:verify}
\begin{tabularx}{\textwidth}{@{}l*{7}{C}c@{}}
\toprule
   & \tabincell{c}{Prediction\\(JIQ)} & \tabincell{c}{$n=500$\\(JIQ)} & \tabincell{c}{$n=1000$\\(JIQ)} & \tabincell{c}{Prediction\\(JIQ-Po$d$)} & \tabincell{c}{$n=500$\\(JIQ-Po$d$)} & \tabincell{c}{$n=1000$\\(JIQ-Po$d$)}\\ 
\midrule
%$p_0$&  $0.4818$&  $0.4834$&  $0.4817$&  $0.4703$&  $0.4739$&  $0.4732$              \\ 
$\hat{q}_{(0,1)}$&  $0.0260$&  $0.0259$&  $0.0261$&  $0.0267$&  $0.0266$&  $0.0266$             \\ 
$\hat{q}_{(0,2)}$&  $0.0269$&  $0.0266$&  $0.0268$&  $0.0281$&  $0.0279$&  $0.0278$              \\ 
$\hat{q}_{(0,3)}$&  $0.0200$&  $0.0200$&  $0.0199$&  $0.0206$&  $0.0203$&  $0.0204$             \\ 
$\hat{q}_{(0,4)}$&  $0.0126$&  $0.0128$&  $0.0127$&  $0.0125$&  $0.0124$&  $0.0125$              \\ 
$\hat{q}_{(0,5)}$&  $0.0072$&  $0.0074$&  $0.0072$&  $0.0067$&  $0.0068$&  $0.0066$             \\  
\addlinespace
\midrule
%$q_0$&  $0.1$&   $0.1003$ & $0.1000$&  $0.1$&   $0.1000$& $0.0999$\\ 
$\hat{q}_1$&  $0.5097$&   $0.5071$ &   $0.5089$ &  $0.5571$&   $0.5523$ &   $0.5532$\\
$\hat{q}_2$&  $0.2210$&    $0.2206$ & $0.2207$&  $0.2931$&    $0.2931$ & $0.2939$\\ 
$\hat{q}_3$&  $0.0958$&    $0.0963$ & $0.0960$&  $0.0487$&    $0.0529$ & $0.0517$\\
$\hat{q}_4$&  $0.0416$     &$0.0424$ & $0.0418$& $0.0010$&    $0.0016$& $0.0014$               \\
$\hat{q}_5$&  $0.0180$&    $0.0186$ & $0.0183$&  $0.0000$&    $0.0000$ & $0.0000$             \\  
\bottomrule
\end{tabularx}
\end{table*}
%\lipsum[2-15]

%Secondly, we check the impact of arrival rate for JIQ and JIQ-Po$d$. According to Figure~\ref{fig:verify20}, we observe that the tail distribution $s_i$ increase with the growth of arrival rate. It also shows that our JIQ-Po$d$ algorithm enjoys lighter $s_i$ than the JIQ algorithm.
%Figure~\ref{fig:compare21} studies the distribution of empty I-queues $p_0$, which also has important impact of system performance.   
%%Figure~\ref{fig:verify22} and Figure~\ref{fig:verify23} study the $p_i$ (a.k.a. the distribution of I-queue queue length) of both algorithms. 
%To our surprise, the $p_0$ of our JIQ and JIQ-Po$d$ is almost the same. This implies that the JIQ-Po$d$ algorithm does not impact I-queue side performance very much. Instead, it optimizes the server side performance directly.
\begin{figure}[tp]
   \centering
\includegraphics[width = 0.48\textwidth]{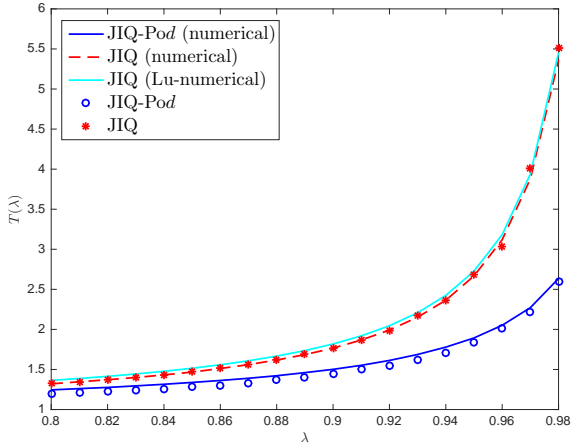}
\vspace{-3mm}
\caption{Response time of JIQ and JIQ-Po$d$ when $N=1000$, $r = 10$ and $\lambda$ changes from $0.9$ to $0.98$.}
\label{fig:verify20}
\vspace{-4mm}
\end{figure}

%Second, we look at the expected task response time. 
Figure~\ref{fig:verify20} shows the task response times of JIQ and JIQ-Po$d$ obtained from prediction and simulation. As we can see, when the server size is $1000$, the maximum relative error is only $3.4 \%$ and the corresponding absolute error is $0.178$. Hence, our theoretical predictions are fairly accurate even for systems of relatively small size.
In Figure~\ref{fig:verify20}, we also compared the prediction of task response times from \cite{Lu11} with ours. The higher arrival rate is, the more accuracy \cite{Lu11} acquires.

\subsection{Impact of ``Delete Request'' Messages}\label{sec:impact}

Recall that in Section~\ref{sec:statement}, we applied a ``delete request'' strategy to the conventional JIQ algorithm (e.g., JIQ-Original) to simplify theoretical analysis.
In the subsection, we explore the impact of such ``delete request'' strategy on our JIQ algorithm  in terms of the communication overhead and mean task response time.
%Such additional ''update'' report happens when an idle server is randomly assigned a task by a scheduler with empty I-queue. 
%Additionally, the ''update'' report also causes some additional communication overheads in the system.
\begin{figure}[tp]
   \centering

\includegraphics[width = 0.48\textwidth]{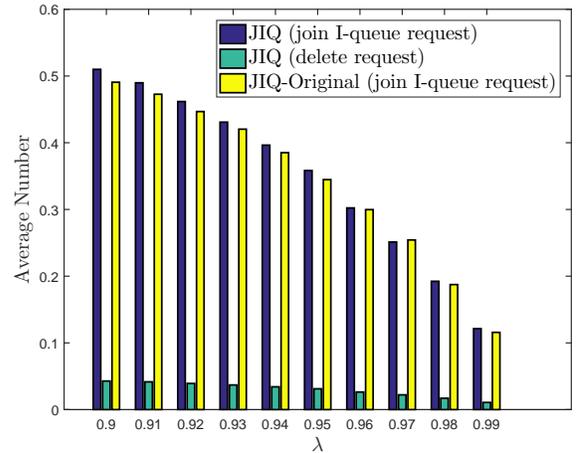}
\vspace{-3mm}
\caption{Average number of ``request'' messages per unit time per server of JIQ and JIQ-Original when $r = 10$ and $\lambda$ changes from $0.9$ to $0.99$.}
\label{fig:compare10}
\vspace{-4mm}
\end{figure}

First, idle servers not only send ``join I-queue request'', but also send ``delete request'' under our JIQ algorithm, which will increase the number of ``request'' messages for each server. Figure~\ref{fig:compare10} studies the average number of ``request'' messages per unit time per server under JIQ-Original and JIQ. It turns out that the ``delete request'' only contributes to a small portion of the overall requests. For instance, when $\lambda = 0.9$, the number of ``delete request'' messages is no more than $8 \%$ of overall requests.  %Besides, the overall communication overhead of JIQ is not increased much compared with that of JIQ-original algorithm.
%Additionally, we also compare performance of JIQ and JIQ-original in terms of $s_i$ and $p_i$in Figure~\ref{fig:compare11}. It turns out that both algorithms' performances are very close to each other, as shown in Figure~\ref{fig:compare11}.

Second, such ``delete request'' strategy has little impact on the mean task response time. Figure~\ref{fig:compare30} compares the mean task response times of different JIQ algorithms. It shows that the mean task response times of both JIQ-Original and JIQ are close to each other. To sum up, the ``delete request'' strategy has little impact on JIQ.%Also, in Figure~\ref{fig:compare31} and Figure~\ref{fig:compare32}, the cumulative distribution of JIQ and JIQ-original are quite close.

\subsection{Comparison of JIQ Algorithms}

Finally, we compare our JIQ-Po$d$ with two other variants, JIQ-Threshold and JIQ-SQ($d$) \cite{Mit16, Lu11}.
\begin{itemize}
\item {\bf{{JIQ-Threshold:}}} There is a threshold $z$ for server queue length. As long as a server has less than or equal to $z$ tasks, it will send a ``join I-queue request'' message to an I-queue. Thus, I-queues contain all servers with less than or equal to $z$ tasks.
\item {\bf{{JIQ-SQ($d$):}}} When an idle server needs to send a ``join I-queue request'' message to an I-queue, it adopts the Po$d$ algorithm to select which I-queue to report.

\end{itemize}

\begin{figure}[tp]
   \centering

\includegraphics[width = 0.48\textwidth]{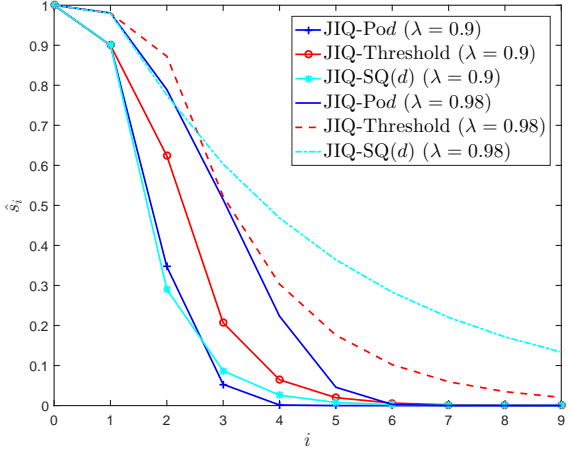}
\vspace{-3mm}
\caption{Tail distribution of three algorithms under light workload ($\lambda=0.9$) and heavy workload ($\lambda=0.98$).}
\label{fig:compare20}
\vspace{-4mm}
\end{figure}

\begin{figure}[tp]
   \centering

\includegraphics[width = 0.48\textwidth]{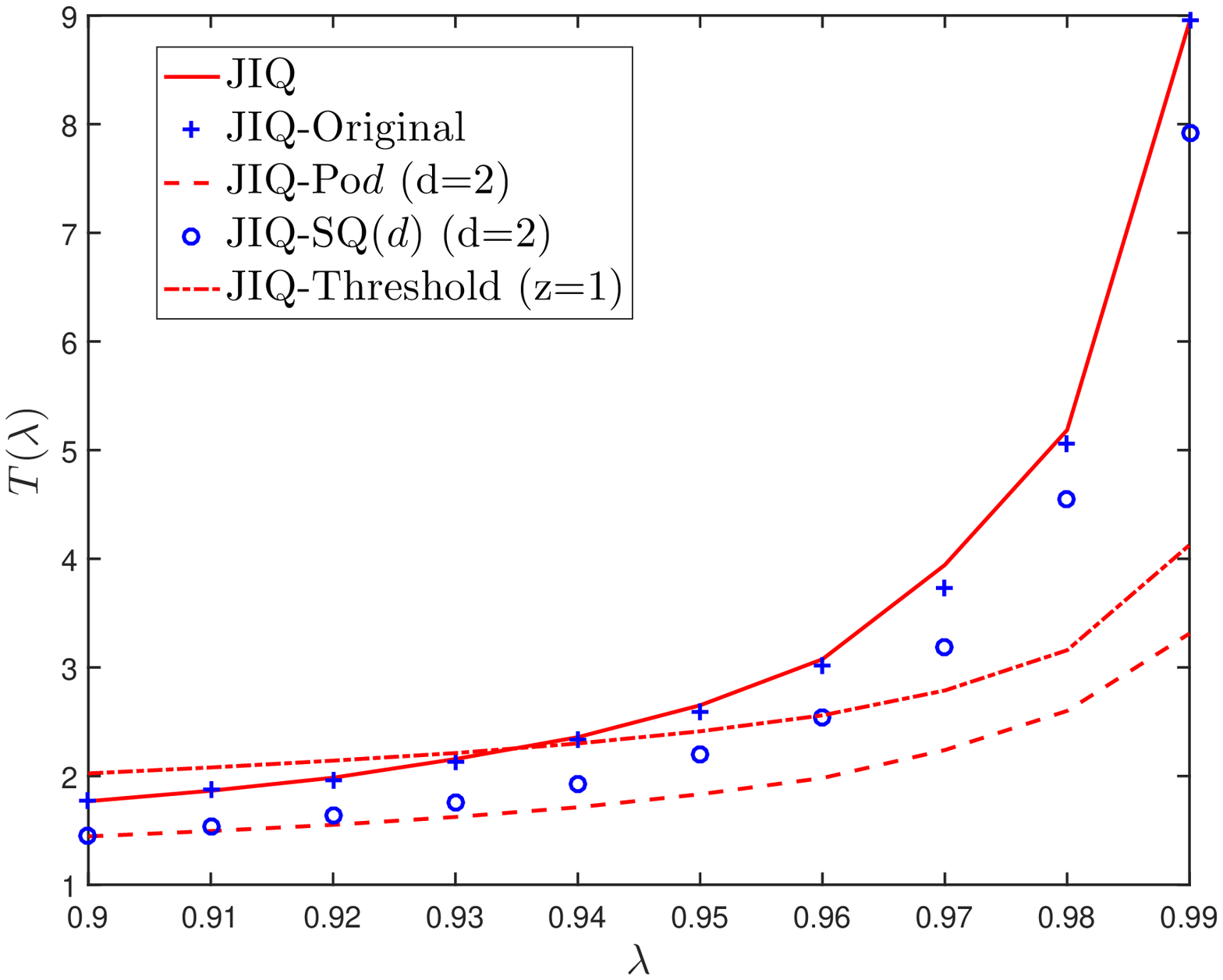}
\vspace{-3mm}
\caption{A comparison of average task response time for five algorithms.}
\label{fig:compare30}
\vspace{-4mm}
\end{figure}

For comparison, we use the tail distribution $\hat{s}_i$ and the mean task response time. Figure~\ref{fig:compare20} compares the tail distributions $\hat{s}_i$ among those three algorithms when $d = 2$ and $z = 1$. In Figure~\ref{fig:compare20}, the JIQ-Po$d$ algorithm always has the lightest tail in the heavy workload region.
Figure~\ref{fig:compare30} compares the mean task response time of different JIQ algorithms. It is shown that the mean task response time of JIQ-Po$d$ is the shortest among five alternative algorithms.
Overall, our JIQ-Po$d$ algorithm achieves the best delay performance compared with other alternative JIQ algorithms.
%\begin{figure}[tp]
%\includegraphics[width = 0.48\textwidth]{JIQ_overhead_distribution.eps}
%\vspace{-3mm}
%\caption{$p_i$ and $s_i$ of JIQ when $\lambda = 0.9$ and $r = 10$.}
%\label{fig:compare11}
%\vspace{-4mm}
%\end{figure}

\begin{figure}[tp]
   \centering

\includegraphics[width = 0.48\textwidth]{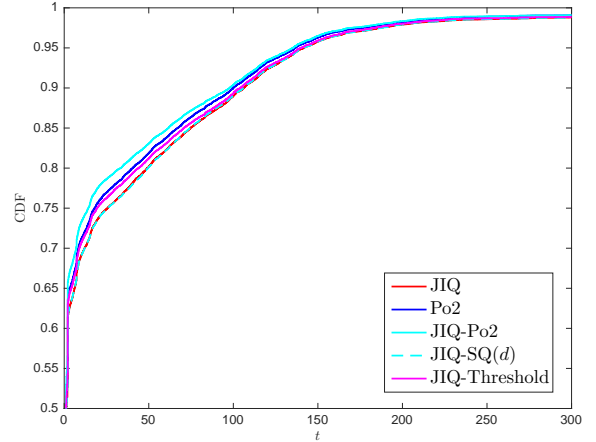}
\vspace{-3mm}
\caption{The cumulative distribution function of task response time when $\lambda = 0.88$, $d = 2$, $r=10$ and threshold is $1$.}
\label{fig:compare60}
\vspace{-4mm}
\end{figure}

\begin{figure}[tp]
   \centering

\includegraphics[width = 0.48\textwidth]{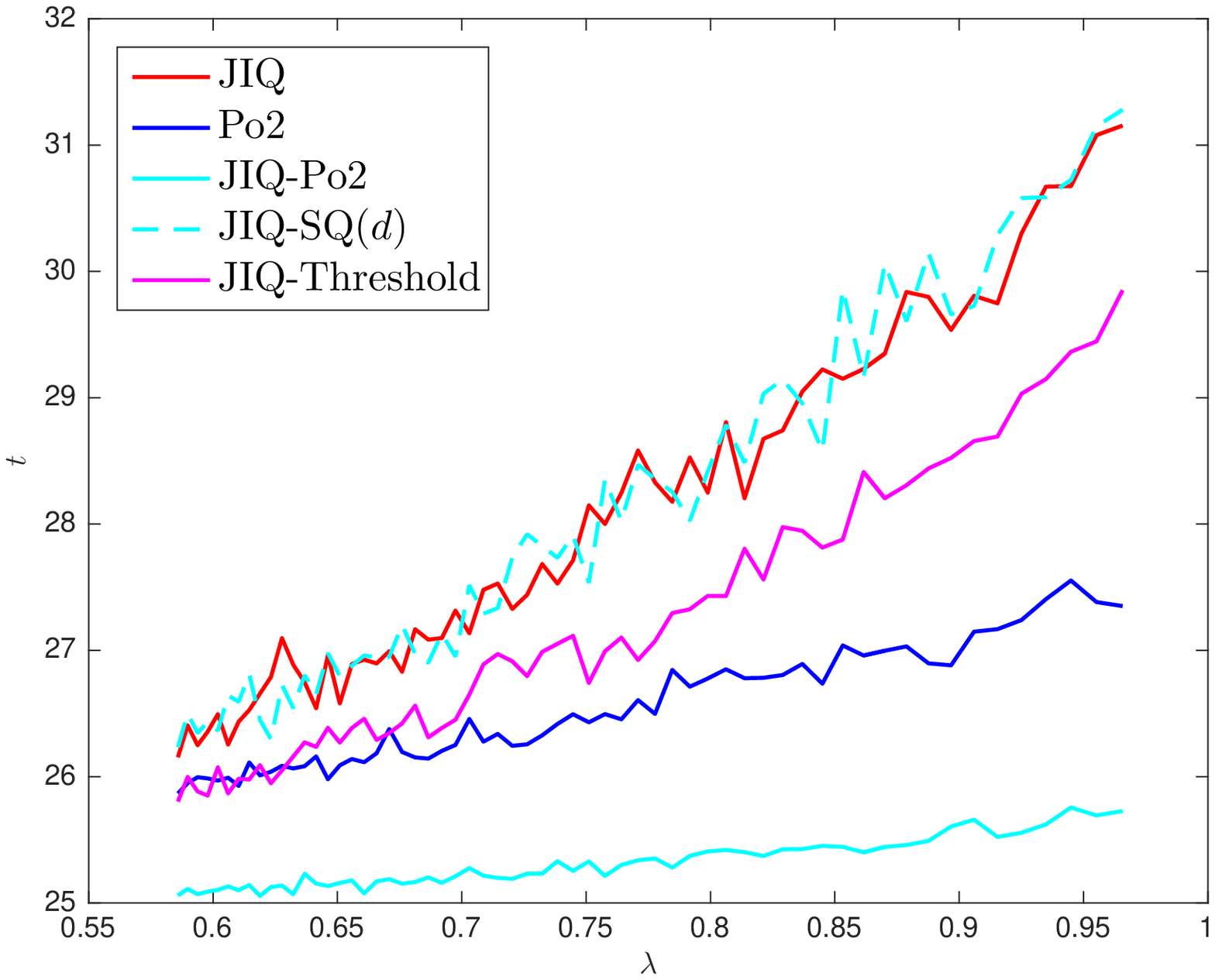}
\vspace{-3mm}
\caption{A comparison of average task response time for five algorithms.}
\label{fig:compare61}
\vspace{-4mm}
\end{figure}

Finally, we conduct trace-driven simulations using real-world data from Google clusters, which contains more than 12,000 tasks over seven hours. In Figure~\ref{fig:compare60}, we compare the cumulative distribution function of task response time among five different algorithms. It shows that the JIQ-Po$2$ algorithm attains the best performance when $\lambda = 0.88$, $d = 2$, $r=10$ (where the threshold is $1$). In Figure~\ref{fig:compare61}, we vary the arrival rate $\lambda$ from $0.58$ to $0.96$. We observe that the JIQ-Po$2$ algorithm achieve the shortest mean task response time and the smallest growth rate.

%\begin{figure}[tp]
%   \centering
%
%\includegraphics[width = 0.48\textwidth]{JIQ_Pod_compare_p0.eps}
%\vspace{-3mm}
%\caption{Empty I-queue fraction $p_0$ of three algorithms when $r = 10$ and $\lambda$ changes from $0.9$ to $0.99$.}
%\label{fig:compare21}
%\vspace{-4mm}
%\end{figure}

%When it comes to heavy workload (e.g., $\lambda=0.98$), the JIQ-Po$d$ algorithm still has the lightest tail but the JIQ-SQ($d$) algorithm has the heaviest tail.

%When it comes to the fraction of empty I-queue, in Figure~\ref{fig:compare21}, it is clear that the JIQ-earlyAssign algorithm always has the fewest empty I-queues while the JIQ-Po$d$ algorithm always has the most. 
%However, this does not imply that JIQ-Po$d$ has the longest mean task response time. On the contrary, the mean task response time of JIQ-Po$d$ is the shortest among five different algorithms in Figure~\ref{fig:compare30}.
%Figure~\ref{fig:compare31} and Figure~\ref{fig:compare32} also show that it performs well in terms of cumulative distribution of task response time whenever under light or heavy workload.

%\begin{figure}[tp]
%\includegraphics[width = 0.48\textwidth]{CDF_task_response_lowRate.eps}
%\vspace{-3mm}
%\caption{Cumulative distribution of task response time when $\lambda=0.9$.}
%\label{fig:compare31}
%\vspace{-4mm}
%\end{figure}
%
%\begin{figure}[tp]
%\includegraphics[width = 0.48\textwidth]{CDF_task_response_highRate.eps}
%\vspace{-3mm}
%\caption{Cumulative distribution of task response time when $\lambda=0.98$.}
%\label{fig:compare32}
%\vspace{-4mm}
%\end{figure} 

\section{Conclusions}
\label{sec:conclusion}

In this paper, we analyzed the JIQ algorithm by the mean-field analysis. 
Then, we proposed a hybrid algorithm called JIQ-Po$d$, which takes the advantage of both JIQ and Po$d$. 
Under the large-system limit, we obtained semi-closed form expressions of the stationary distributions of JIQ and JIQ-Po$d$.
Our theoretical results fit the simulation results well in reasonable large systems, e.g., $N=1000$. 
In addition, our simulation results show that our JIQ-Po$d$ outperforms many JIQ variants in all conditions.

\bibliographystyle{IEEEtran}

\appendix
\label{sec:appendix}

\subsection{Proof of Theorem~\ref{thm:positive-recurrent}}
\label{apx-a}

Set the potential equation ${V}({\mathbf{z}})$ as
\begin{equation}\label{potential-equation}
{V}({\mathbf{z}}) = \sum\limits_{i = 1}^N {z_{i,1}^2}  + \sum\limits_{i = 1}^N {z_{i,2}^2}
\end{equation}
where ${\mathbf{z}} = \left\{ {\left( {z_{i,1}^{(N)}(t),z_{i,2}^{(N)}(t)} \right)} \right\}_{i = 1}^N$.

Let ${{q_{{\mathbf{z}},{\mathbf{w}}}}}$ be the transition rate from system state $\mathbf{z}$ to $\mathbf{w}$. The system state changes only when there is a task-arrival or a task-departure event happens.
% For the system with state $\left\{ {{{\mathbf{x}}_{\mathbf{Q}}},{{\mathbf{x}}_{\mathbf{P}}}} \right\}$, the next possible state with a task arrival or departure can only be one of the following four: $\left\{ {{{\mathbf{x}}_{\mathbf{Q}}} + {e_i},{{\mathbf{x}}_{\mathbf{P}}} - {e_j}} \right\}$, $\left\{ {{{\mathbf{x}}_{\mathbf{Q}}} + {e_i},{{\mathbf{x}}_{\mathbf{P}}}} \right\}$, $\left\{ {{{\mathbf{x}}_{\mathbf{Q}}} - {e_i},{{\mathbf{x}}_{\mathbf{P}}}} \right\}$ and $\left\{ {{{\mathbf{x}}_{\mathbf{Q}}} - {e_i},{{\mathbf{x}}_{\mathbf{P}}} + {e_j}} \right\}$.
We consider the Lyapunov drift as follows:
\begin{equation}\label{positive-recurrent-drift}
\begin{split}
& \sum\limits_{{\mathbf{w}} \ne {\mathbf{z}}} {{q_{{\mathbf{z}},{\mathbf{w}}}}\left[ {{V}({\mathbf{w}}) - {V}({\mathbf{z}})} \right]} \\
&  = \sum\limits_{{\mathbf{w}} \ne {\mathbf{z}}} {{q_{{\mathbf{z}},{\mathbf{w}}}}\left[ {\sum\limits_{i = 1}^N {(w_{i,1}^2 - z_{i,1}^2)}  + \sum\limits_{i = 1}^N {(w_{i,2}^2 - z_{i,2}^2)} } \right].} 
\end{split}
\end{equation}
By the Foster-Lyapunov theorem, we only need to show that for any fixed $N$ and $M$,
\begin{equation}\label{positive-recurrent-target}
\begin{split}
& \sum\limits_{{\mathbf{w}} \ne {\mathbf{z}}} {{q_{{\mathbf{z}},{\mathbf{w}}}}\left[ {{V}({\mathbf{w}}) - {V}({\mathbf{z}})} \right]} \\
&  \le 2(\lambda  - 1)\sum\limits_{i = 1}^N {{z_{i,1}}}  + (1 + {M^2} + \lambda )N 
\end{split}
\end{equation}
This is because: 
\begin{enumerate}

\item If $\sum\limits_{i = 1}^{N} {{z_{i,1}}}  > \frac{{(1 + {M^2} + \lambda )N}}{{2(1 - \lambda )}}$, we have

\[
\begin{split}
& \sum\limits_{{\mathbf{w}} \ne {\mathbf{z}}} {{q_{{\mathbf{z}},{\mathbf{w}}}}\left[ {{V}({\mathbf{w}}) - {V}({\mathbf{z}})} \right]} < 0.\hfill \\ 
\end{split} 
\]

\item If $\sum\limits_{i = 1}^{N} {{z_{i,1}}}  \leq \frac{{(1 + {M^2} + \lambda )N}}{{2(1 - \lambda )}}$, we have
\[
\begin{split}
& \sum\limits_{{\mathbf{w}} \ne {\mathbf{z}}} {{q_{{\mathbf{z}},{\mathbf{w}}}}\left[ {{V}({\mathbf{w}}) - {V}({\mathbf{z}})} \right]} < \infty.\hfill \\ 
\end{split} 
\]

\end{enumerate}

In terms of ${z_{i,2}}$, it increases from $0$ to a positive number in $[1,M]$ when the $i$th server becomes idle. In other cases, ${z_{i,2}}$ remains unchanged or decreases to $0$.
As ${z_{i,2}} \in [0,M]$ and the processing rate for each server is $1$, we obtain 

\begin{equation}\label{drift-part-1}
\sum\limits_{{\mathbf{w}} \ne {\mathbf{z}}} {{q_{{\mathbf{z}},{\mathbf{w}}}}\left[ {\sum\limits_{i = 1}^N {(w_{i,2}^2 - z_{i,2}^2)} } \right]}  \le N({M^2} - {0^2}) = N{M^2}.
\end{equation}

In terms of ${z_{i,1}}$, it increases by $1$ when a task arrives to the $i$th server; it decrease by $1$ when a task departures the $i$th server. Let ${p_0^{(N)}}$ be the fraction of empty I-queues. Recall the evolution of a JIQ system, when a new task arrives, we obtain
\[ \Pr \{ \mbox{meet a non-empty I-queue} \} = 1-{p_0^{(N)}} \]
and
\[ \Pr \{ \mbox{meet an empty I-queue} \} = {p_0^{(N)}}. \]
For task-arrival events, we have
\[
\begin{array}{l}
\sum\limits_{{\mathbf{w}} \ne {\mathbf{z}}} {q_{{\mathbf{z}},{\mathbf{w}}}^{\mbox{(arrival)}}\left[ {\sum\limits_{i = 1}^N {(w_{i,1}^2 - z_{i,1}^2)} } \right]} \\
 \le \lambda Np_0^{(N)}\sum\limits_{i = 1}^N {\frac{{{{({z_{i,1}} - 1)}^2} - z_{i,1}^2}}{N}}  + \lambda N(1 - p_0^{(N)})({1^2} - {0^2})\\
 \le 2\lambda \sum\limits_{i = 1}^N {{z_{i,1}}}  + \lambda N.
\end{array}
\]
For task-departure events, recall that the server processing rate is $1$, we have
\[
\begin{array}{l}
\sum\limits_{{\mathbf{w}} \ne {\mathbf{z}}} {q_{{\mathbf{z}},{\mathbf{w}}}^{\mbox{(departure)}}\left[ {\sum\limits_{i = 1}^N {(w_{i,1}^2 - z_{i,1}^2)} } \right]} \\
 \le \sum\limits_{i = 1}^N {\left[ {{{({z_{i,1}} + 1)}^2} - z_{i,1}^2} \right]} \\
 =  - 2\sum\limits_{i = 1}^N {{z_{i,1}}}  + N.
\end{array}
\]
Thus, we have 
\begin{equation}\label{drift-part-2}
\sum\limits_{{\mathbf{w}} \ne {\mathbf{z}}} {{q_{{\mathbf{z}},{\mathbf{w}}}}\left[ {\sum\limits_{i = 1}^N {(w_{i,1}^2 - z_{i,1}^2)} } \right]}  \le 2(\lambda  - 1)\sum\limits_{i = 1}^N {{z_{i,1}}}  + (1 + \lambda )N.
\end{equation}

Finally, we sum up ~\eqref{drift-part-1} and ~\eqref{drift-part-2} to have ~\eqref{positive-recurrent-target}.  
This completes the proof.

\subsection{$f(p_0)$ is differentiable and monotonically increasing}
\label{apx-b}

First of all, we will show that $f(p_0)$ is differentiable over the interval $(0, 1)$. According to \eqref{eq:gamma},
it suffices to prove that
\[
h(p_0) \triangleq {\Gamma\left(\frac{{r + {p_0}}}{{{p_0}}}\right) - \Gamma\left(\frac{{r + {p_0}}}{{{p_0}}},r\left( - \lambda  + \frac{1}{{{p_0}}}\right)\right)} 
\]
is differentiable over $(0, 1)$. Recall that $h(p_0)$ can be rewritten as
\[
h(p_0) = \int_{0}^{a(p_0)} g(p_0, t) dt
\]
where $a(p_0) = r\left( - \lambda + \frac{1}{p_0} \right)$ and $g(p_0, t) = t^{\frac{r}{p_0}} e^{-t}$.
By the Leibniz's integral rule, $h(p_0)$ is differentiable as long as 
\begin{itemize}
    \item $a(p_0)$ has continuous derivative over $(0, 1)$;
    \item $g(p_0, t)$ and its partial derivative $\frac{\partial}{\partial p_0} g(p_0, t)$ are continuous in the region 
    of $0 < p_0 < 1$ and $0 \le t \le a(p_0)$.
\end{itemize}
Both requirements can be easily verified.

Next, we will show that $f(p_0)$ is monotonically increasing over $(0, 1)$. It suffices to show that
$f'(p_0) > 0$ over $(0,1)$, since $f(p_0)$ is differentiable. To this end, we define 
$g_i(p_0)$ as
\begin{equation}\label{I-queue-steady}
{g_i}({p_0}) \triangleq \left\{ \begin{array}{l}
{p_0},i = 0,\\
\frac{{{r^i}{{(1 - \lambda {p_0})}^i}}}{{\prod\limits_{j = 1}^i {(r + j{p_0})} }}{p_0},i \ge 1.
\end{array} \right.
\end{equation}
Clearly, we have $f(p_0) = \sum\limits_{i = 0}^\infty  {{g_i}({p_0})}$. Hence, we need to show that 
$\sum\limits_{i = 0}^\infty  {{g'_i}({p_0})} > 0$. A key observation is the following. 

\begin{lemma}\label{lem:sign}
If there exists an integer
$k$ such that $g'_k(p_0) < 0$, then for all $i > k$ we have $g'_i(p_0) < 0$.
\end{lemma}
\begin{IEEEproof}
By ~\eqref{I-queue-steady}, we have 
\begin{equation}\label{eq:11}
g_{i+1}(p_0) = \frac{{r(1 - \lambda {p_0})}}{r + (i+1){p_0}}g_{i}(p_0), \mbox{ for } i \geq 0.
\end{equation}
 
Taking derivatives on both sides, we obtain
 \begin{equation}\label{1}
  g'_{i+1}(p_0) =  - \frac{{r(r\lambda  + (i+1))}}{{{{(r + (i+1){p_0})}^2}}}g_{i}(p_0) + \frac{{r(1 - \lambda {p_0})}}{{r + (i+1){p_0}}} g'_{i}(p_0).
 \end{equation}

Note that $1 - \lambda {p_0} > 0$. Hence, if $g'_k(p_0) < 0$, then we have $g'_{k+1}(p_0) < 0$.
\end{IEEEproof}

By Lemma~\ref{lem:sign}, we only need to consider two cases:
\begin{enumerate}
    \item For all $i \ge 1$, $g'_i(p_0) \ge 0$.
    \item There exists an integer
$k$ such that for all $i < k$, $g'_i(p_0) \ge 0$ and for all $i \ge k$, $g'_i(p_0) < 0$.
\end{enumerate}

For Case~1), we have $\sum\limits_{i = 0}^\infty  {{g'_i}({p_0})} > 0$, because $g'_0(p_0) = 1$ and 
${g'_i}({p_0}) \ge 0$ when $i \ge 1$.

For Case~2), we need some additional argument. By \eqref{eq:11}, we have
\begin{equation*}
(r + (i+1){p_0}) g_{i+1}(p_0) = r(1 - \lambda {p_0}) g_{i}(p_0), \mbox{ for } i \geq 0.
\end{equation*}
Hence,
\[
\sum_{i = 1}^\infty (r + i{p_0}) g_{i}(p_0) = r(1 - \lambda {p_0}) \sum_{i = 0}^\infty g_{i}(p_0).
\]
It follows that
 \begin{equation*}
  \sum\limits_{i = 0}^\infty  {i{g_i}(p_0)}  = r - r\lambda \sum\limits_{i = 0}^\infty  {{g_i}(p_0)}
 \end{equation*}
and
 \begin{equation*}
  \sum\limits_{i = 0}^\infty  {i{g'_i}(p_0)}  =  - r\lambda \sum\limits_{i = 0}^\infty  {{g'_i}(p_0)}.
 \end{equation*}

Recall that $k$ is the smallest integer such that $g'_k(p_0) < 0$. Thus, we have 
\[
\sum\limits_{i = 0}^\infty  k g'_i(p_0)  > \sum\limits_{i = 0}^\infty  i g'_i(p_0).
\]
Therefore, 
\[
\sum\limits_{i = 0}^\infty  k g'_i(p_0)  > - r\lambda \sum\limits_{i = 0}^\infty  {{g'_i}(p_0)}.
\]
This implies that $\sum\limits_{i = 0}^\infty  {{g'_i}({p_0})} > 0$.

% Generated by IEEEtran.bst, version: 1.13 (2008/09/30)

\end{document}